\documentclass{PoS}

\def\a{{\alpha}}

\def\d{{\delta}}
\def\D{{\Delta}}

\def\g{{\gamma}}

\def\O{{\Omega}}

\def\s{{\sigma}}

\def\tr{\text{tr}}

\def\mc#1{{\mathcal #1}}

\def\eqref#1{{(\ref{#1})}}

\def\tr{\mathrm{tr}}

\title{Towards a direct lattice calculation of $m_d - m_u$ }

\ShortTitle{$m_d - m_u$}

\author{\speaker{Andr\'{e} Walker-Loud}\thanks{I am indebted to my collaborators on this project, Christopher Aubin, Carl Carlson, Will Detmold, Kostas Orginos and Brian Tiburzi.}\\
        College of William and Mary\\
        Lawrence Berkeley National Laboratory\\
        E-mail: \email{awalker-loud@lbl.gov}}

\abstract{We describe an independent method for determining the strong-isospin breaking mass parameter, $2\d = m_d - m_u$, which utilizes the baryon spectrum.  We use a prudent partially quenched choice of splitting the valence quark masses symmetrically about the light sea quark mass.  This choice has the consequence of mitigating the most severe partial quenching artifacts.
We also discuss the most significant hurdle to this method which is determining the electromagnetic self-energy of the neutron-proton mass splitting, a challenge which lacks a satisfactory answer.  Despite these issues, the phenomenologically interesting dependence of $m_n - m_p$ on $\d$ can be determined.}

\FullConference{The XXVIII International Symposium on Lattice Field Theory, Lattice2010\\
		June 14-19, 2010\\
		Villasimius, Italy}

\begin{document}

%
\section{Motivation}
The quark masses are fundamental parameters of the QCD Lagrangian, inherited from their interactions with the electro-weak sector of the standard model.  Determining these fundamental parameters remains an interesting topic for lattice QCD.  There are now several groups reporting values of the quark masses to high precision.  We present a new method for determining the strong isospin breaking mass parameter, $2\d = m_d- m_u$.  More than being interesting in its own right, we would like to know how the neutron-proton mass splitting depends upon this parameter; while the universe largely respects isospin symmetry, there are few important phenomenological quantities which have a strong dependence on $m_n - m_p$, \textit{eg.} the neutron lifetime, charge symmetry breaking~\cite{Miller:2006tv} and time-reversal violating pion-nucleon interactions~\cite{Mereghetti:2010tp} to name a few.

We begin in Sec.~\ref{sec:EMselfenergy} by detailing the most significant systematic, the hadron electromagnetic (EM) self-energy.  In Sec.~\ref{sec:PQ}, we describe our prudent partially quenched setup, placing the isospin breaking in the valence sector only, but split symmetrically about the sea quark mass.  In Sec.~\ref{sec:results} we display the results of our calculation and in Sec.~\ref{sec:conclusions} we conclude.

%
\section{Electromagnetic Self Energy\label{sec:EMselfenergy}}
Understanding hadron electromagnetic self energies is an old problem, originally motivated to explain the proton-neutron mass splitting~\cite{Zee:1971df}.  The EM correction presents the most significant challenge in determining the strong isospin breaking parameter $2\d = m_d - m_u$ from the nucleon spectrum.
Using the Cottingham formula, a hadrons electromagnetic self energy can be determined from knowledge of its electromagnetic structure~\cite{Cottingham:1963zz}.  For example, for spin $1/2$ baryons we have
\begin{equation}
\Delta M_\g^{B} = -\frac{\alpha_{f.s.}}{4\pi^2} \int_0^\infty dQ^2 \int_{-Q}^{Q} d\nu
	\frac{\sqrt{Q^2 - \nu^2}}{Q^2} \left[
	3W_1(Q^2,i\nu)
	-\left( 1 - \frac{\nu^2}{Q^2} \right) W_2(Q^2,i\nu) \right]\, .
\end{equation}
A dispersion relation can be used to determine the structure functions $W_i(Q^2, i\nu)$, allowing for a model independent determination of the EM self-energy.  There are two difficulties with this approach however.  First, it has been known since 1966 that a subtracted dispersion relation is required for $W_1(Q^2,i\nu)$~\cite{Harari:1966mu}.  Second, this integral requires renormalization.  Fortunately, the second of these issues was solved by Collins~\cite{Collins:1978hi} who used the operator product expansion to perform the renormalization and connect the Cottingham formula to the QCD Lagrangian.%
\footnote{We are indebted to L.~Yaffe for bring this work to our attention.}
Unfortunately, there is not yet a satisfying resolution of the first issue dealing with the subtracted dispersion relation.

In Ref.~\cite{Gasser:1982ap}, Gasser and Leutwyler provided a comprehensive determination of the quark mass parameters of the QCD Lagrangian.  One method relies on the baryon spectrum, which must be corrected for EM self-energies.
In their work, they acknowledged the need for a subtracted dispersion relation but proceeded to ignore this issue, as the subtraction constant could not be computed.  In the case of the spin $1/2$ baryons, they used the elastic scattering data to determine the bulk of the EM self-energies and knowledge at the time of the inelastic scattering data to estimate the uncertainties.  In Table~\ref{tab:GL_EM}, we collect these results for the nucleon and cascade.  If one uses a subtracted dispersion relation, then the elastic contribution becomes $\D M^\g_{n-p} = -1.39$~\texttt{MeV} plus the subtraction term.  The computable elastic contribution is nearly a factor of two different from that in Ref.~\cite{Gasser:1982ap}, indicating the possibility of a large systematic not accounted for.
With modern knowledge of nucleon electromagnetic structure, there is hope to be able to update the Gasser and Leutwyler determination to properly account for the required subtraction~\cite{EMselfenergy}.
\begin{table}
\center
\begin{tabular}{c|ccc}
\hline\hline
$m_B - m_{B^\prime}$& Experiment& $\D M_\g$& QCD \\
\hline
$m_{\Xi^-} - m_{\Xi^0} \texttt{ [MeV]}$& $6.85\pm0.21$& $0.86\pm0.30$& $5.99\pm0.37$\\
$m_n - m_p \texttt{ [MeV]}$& 1.29& $-0.76\pm0.30$& $2.05\pm0.30$ \\
\hline\hline
\end{tabular}
\caption{\label{tab:GL_EM} Estimates from Ref.~\cite{Gasser:1982ap} of the EM self-energy.}
\end{table}

%
\section{Partially Quenched Set Up\label{sec:PQ}}
Presently, lattice ensembles are all generated in the isospin symmetric limit with $m_d = m_u = \hat{m}$.  However, as is known, one can still study isospin breaking corrections by including $m_d \neq m_u$ effects in the valence sector only using a partially quenched framework~\cite{Beane:2002vq,Tiburzi:2005na,Beane:2006fk}.
In this work, we use the prudent choice~\cite{WalkerLoud:2009nf}
\begin{equation}
m_u^{valence} = \hat{m} - \d\, ,
\qquad
m_d^{valence} = \hat{m} + \d\, .
\end{equation}
The $SU(4|2)$ partially quenched heavy baryon Lagrangian, needed to quantify the systematics, can be found in Refs.~\cite{Beane:2002vq,Tiburzi:2005na,WalkerLoud:2009nf}.  We use the conventions of Ref.~\cite{WalkerLoud:2009nf}, where it was demonstrated that with our particular choice of quark masses, the nucleon masses are given through NLO by%
\footnote{We have defined $\hat{\d} \equiv 2 B \d$ where $B = -\langle \bar{q} q \rangle / f^2$ is the chiral condensate.}
\begin{eqnarray}\label{eq:mNnlo}
m_{n,p} &=& M_0 
	\pm \frac{\hat{\d}}{(4\pi f_\pi)} \frac{\a_N}{2}
	+ \frac{m_\pi^2}{(4\pi f_\pi)} \left(\frac{\a_N}{2} + \s_N^r(\mu)\right)
	-3 g_A^2 \frac{\mc{F}(m_\pi,0,\mu)}{(4\pi f_\pi)^2}
	-\frac{8g_{\D N}^2}{3}\frac{\mc{F}(m_\pi,\D,\mu)}{(4\pi f_\pi)^2}
\nonumber\\&& +\frac{3\pi \D_{PQ}^4(g_A + g_1)^2}{8m_\pi (4\pi f_\pi)^2}\, ,
\end{eqnarray}
where $\mc{F}(m_\pi,\D,\mu)$ can be found in Refs.~\cite{Tiburzi:2005na,WalkerLoud:2009nf} and $\mc{F}(m_\pi,0,\mu) = \pi m_\pi^3$.
In Eq.~\eqref{eq:mNnlo}, the first line (of both $m_p$ and $m_n$) is exactly the same as the contributions from $SU(2)$, including isospin breaking; the second line is the contribution from partial quenching.  In our construction, the isospin breaking mass parameter in the valence sector also describes the partial quenching,
\begin{equation}
\D_{PQ}^2 = \hat{\d}\, .
\end{equation}
We have isolated these terms to remind the reader they are unphysical partial quenching artifacts.

In the proton-neutron mass splitting, with our choice of partial quenching, the NLO contributions to the masses exactly cancel.  This is not true for general choices of partial quenching, and one of the great benefits to our choice.  Further, it is the NLO terms in the nucleon mass which render the expansion difficult~\cite{WalkerLoud:2008bp,WalkerLoud:2008pj}, and therefore this symmetric splitting of the valence quark masses about the degenerate sea mass renders the chiral expansion for $m_n - m_p$ as well behaved as that for pions;
\begin{equation}\label{eq:MsplitLO}
m_n - m_p = \frac{\a_N}{(4\pi f_\pi)} \hat{\d}
	+\mc{O}(\hat{\d}^2, \hat{\d}m_\pi^2)\, .
\end{equation}

At next-to-next-to leading order (NNLO), this partially quenched construction incurs its first error.  The $SU(2)$ NNLO quark mass Lagrangian contains nine operators~\cite{Tiburzi:2005na}.  With our symmetric choice of giving the valence quark masses isospin breaking, there are two operators which do not give the correct mass shift,
\begin{equation}
\mc{L}_M \supset \frac{1}{(4\pi f)^3} \left\{
	b_5^M \bar{N} N \, \tr ( \mc{M}_+^2 )
	+ t_3^M  \, \bar{T}^\mu T_\mu \tr (\mc{M}_+^2)
	\right\}\, .
\end{equation}
With our partially quenched $SU(4|2)$ choice, we find ($SU(2)$ denotes the correct mass shift)
\begin{equation}
SU(4|2): 
\begin{array}{c}
\d m_N = \frac{b_5^M(m_\pi^4)}{2(4\pi f_\pi)^3}
\\ \\
\d m_\D = \frac{t_3^M(m_\pi^4)}{2(4\pi f_\pi)^3}
\end{array}\, ,
\qquad
SU(2):
\begin{array}{c}
\d m_N = \frac{b_5^M(m_\pi^4 + \hat{\d}^2)}{2(4\pi f_\pi)^3}
\\ \\
\d m_\D = \frac{t_3^M(m_\pi^4 + \hat{\d}^2)}{2(4\pi f_\pi)^3}
\end{array}\, .
\end{equation}
But these errors are isoscalar and so in the mass splittings, they exactly cancel.  At this order in the expansion, we have introduced no uncontrolled errors.  The full mass splitting to NNLO is given by
\begin{eqnarray}\label{eq:MsplitNNLO}
m_n - m_p = \frac{\hat{\d}}{(4\pi f_\pi)} \bigg\{&&
	 \a_N
	+\frac{m_\pi^2}{(4\pi f_\pi)^2} (b_1^M + b_6^M)
	+\frac{\mc{J}(m_\pi,\D,\mu)}{(4\pi f_\pi)^2} 4g_{\D N}^2 \left( \frac{5}{9}\g_N - \a_N \right)
\nonumber\\&&
	+ \frac{m_\pi^2}{(4\pi f_\pi)^2}\bigg[ \frac{20}{9}\g_N g_{\D N}^2 - 4\a_N (g_A^2 +g_{\D N}^2)
	-\a_N(6g_A^2 + 1) \ln \left( \frac{m_\pi^2}{\mu^2} \right) \bigg]
\nonumber\\&&
	+\frac{\a_N\D_{PQ}^4}{m_\pi^2(4\pi f_\pi)^2} \left( 2 - \frac{3}{2}(g_A + g_1)^2 \right)
\bigg\}\, ,
\end{eqnarray}
where the function $\mc{J}(m_\pi,\D,\mu)$ can be found in Refs.~\cite{Tiburzi:2005na,WalkerLoud:2009nf}.
The first two lines are the expected contribution from $SU(2)$ while the last line is the unphysical effects from partial quenching.

%
\section{Numerical Results\label{sec:results}}
Given the difficulties discussed in Sec.~\ref{sec:EMselfenergy}, there are two ways to proceed.  If a suitable determination of the subtraction constant in the nucleon EM self-energy can not be determined, then the lattice computation can be used, in conjunction with the experimental neutron-proton mass splitting to determine the subtraction constant.  Up to $SU(3)$ and large $N_c$ corrections, this is the same subtraction constant which appears in the $m_{\Xi^-} - m_{\Xi^0}$ EM self-energy.  So the cascade mass splitting can then be postdicted as a consistency check.  However, if the subtraction constant can be determined, then we have an independent means of determining the fundamental parameter $2\d = m_d - m_u$.  We will assume the latter to be the case and proceed.

We shall use our lattice calculations of the cascade spectrum to determine $\d$.  This will allow for a postdiction of $m_n - m_p$ which is phenomenologically more important.  Further, our numerical cascade correlation functions are significantly less noisy, allowing for a more precise determination.  Until we have a better determination of the EM self energy~\cite{EMselfenergy}, we shall use the corrections determined by Gasser and Leutwyler~\cite{Gasser:1982ap} (which ignore the subtraction constant), given in Table~\ref{tab:GL_EM}.  Similar to the nucleons, the cascades are an isospin $1/2$ multiplet with a low lying spin and isospin $3/2$ resonance.  Therefore, the $SU(2)$ heavy-baryon Lagrangian for the $\Xi$s will be the same as that for the nucleons, differing only the value of the low-energy constants~\cite{Tiburzi:2008bk}.

\begin{table}
\begin{tabular}{cccc|cc|cccc}
\hline\hline
\multicolumn{4}{c|}{ensemble: $a_t m_s = -0.0743$}& $m_\pi$& $m_K$& \multicolumn{4}{c}{$a_t\d\ [N_{cfg}\times N_{src}]$ } \\
\hline
$L$& $T$& $a_t m_l$& $a_t m_l^{val}$& $[\texttt{MeV}]$& $[\texttt{MeV}]$& 0.0002& 0.0004& 0.0010& 0.0020 \\
\hline
16& 128& -0.0830& -0.0830& 490& 630&$167\times 25$&--&$167\times 25$&-- \\
16& 128& -0.0840& -0.0840& 420& 592&$166\times 25$&$166\times 25$&$166\times 25$&$166\times 50$  \\
20& 128& -0.0840& -0.0840& 420& 592&$120 \times 25$&--&-- &--\\
24& 128& -0.0840& -0.0840& 420& 592&$97 \times 25$&$100\times10$&$193 \times 25$&--\\
24& 128& -0.0860& -0.0860& 244& 506&$108\times 25$&--&--&-- \\
32& 256& -0.0860& -0.0860& 244& 506&$104 \times 7$&--&--&-- \\
\hline\hline
\end{tabular}
\caption{\label{tab:lattparams}Parameters used in this work.}
\end{table}
For this work, we chose to use the tree level clover improved anisotropic lattices generated by the Spectrum Collaboration~\cite{Lin:2008pr}.  In Table~\ref{tab:lattparams}, we list the values of the various lattice parameters used in this work.  To generate the relatively high statistics, we made use of the EigCG deflation algorithm~\cite{Stathopoulos:2007zi}.
While we have results at several values of $\d$, $m_\pi$ and the volume, we are limited to a single lattice spacing.
Because we are focussed on mass splittings, the leading discretization effects exactly cancel, but this is a systematic we can not control on this ensemble.
We have used $m_\O$ to set the scale, finding $a_t\simeq 0.0324$~\texttt{fm} ($a_s \simeq 0.113$~\texttt{fm}).  

\begin{figure}
\begin{tabular}{cc}
\includegraphics[width=0.5\textwidth]{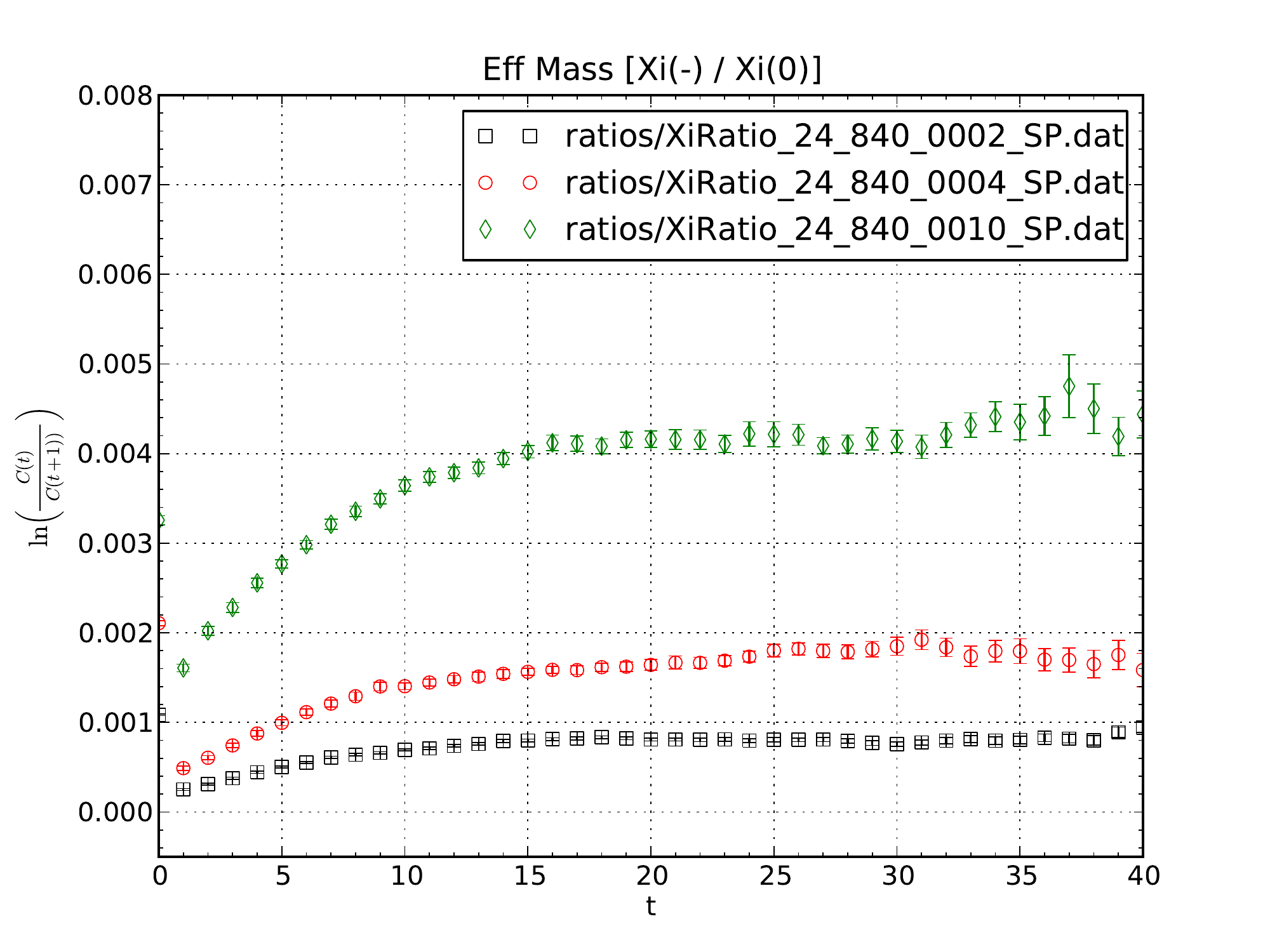}
&
\includegraphics[width=0.5\textwidth]{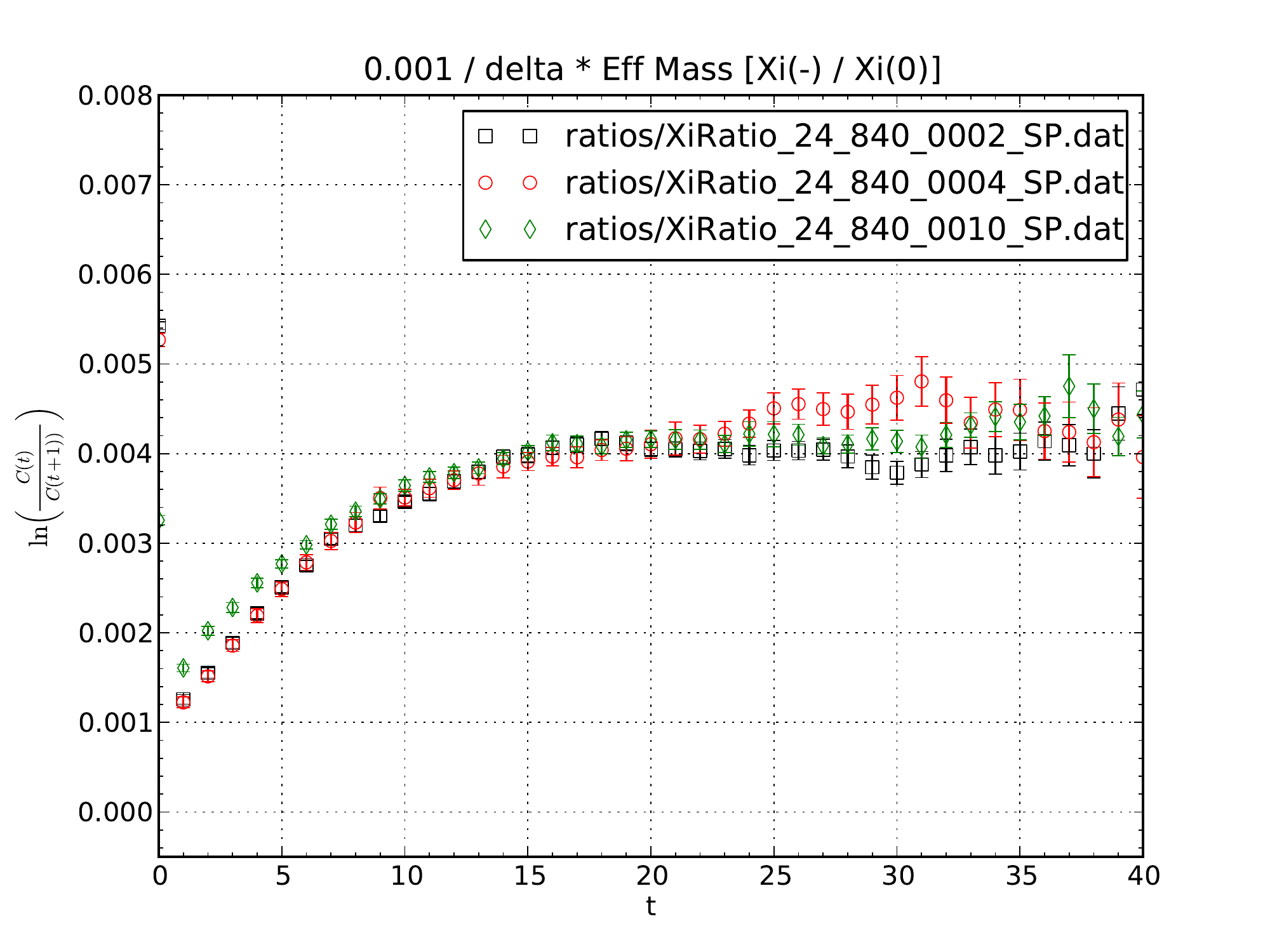}
\end{tabular}
\caption{\label{fig:xiRatio} Effective mass plots of the ratio correlators $\Xi^-(t) / \Xi^0(t)$.  In the right panel, we plot the effective masses scaled by $0.001 / \d$.}
\end{figure}

\begin{figure}
\begin{tabular}{cc}
\includegraphics[width=0.5\textwidth]{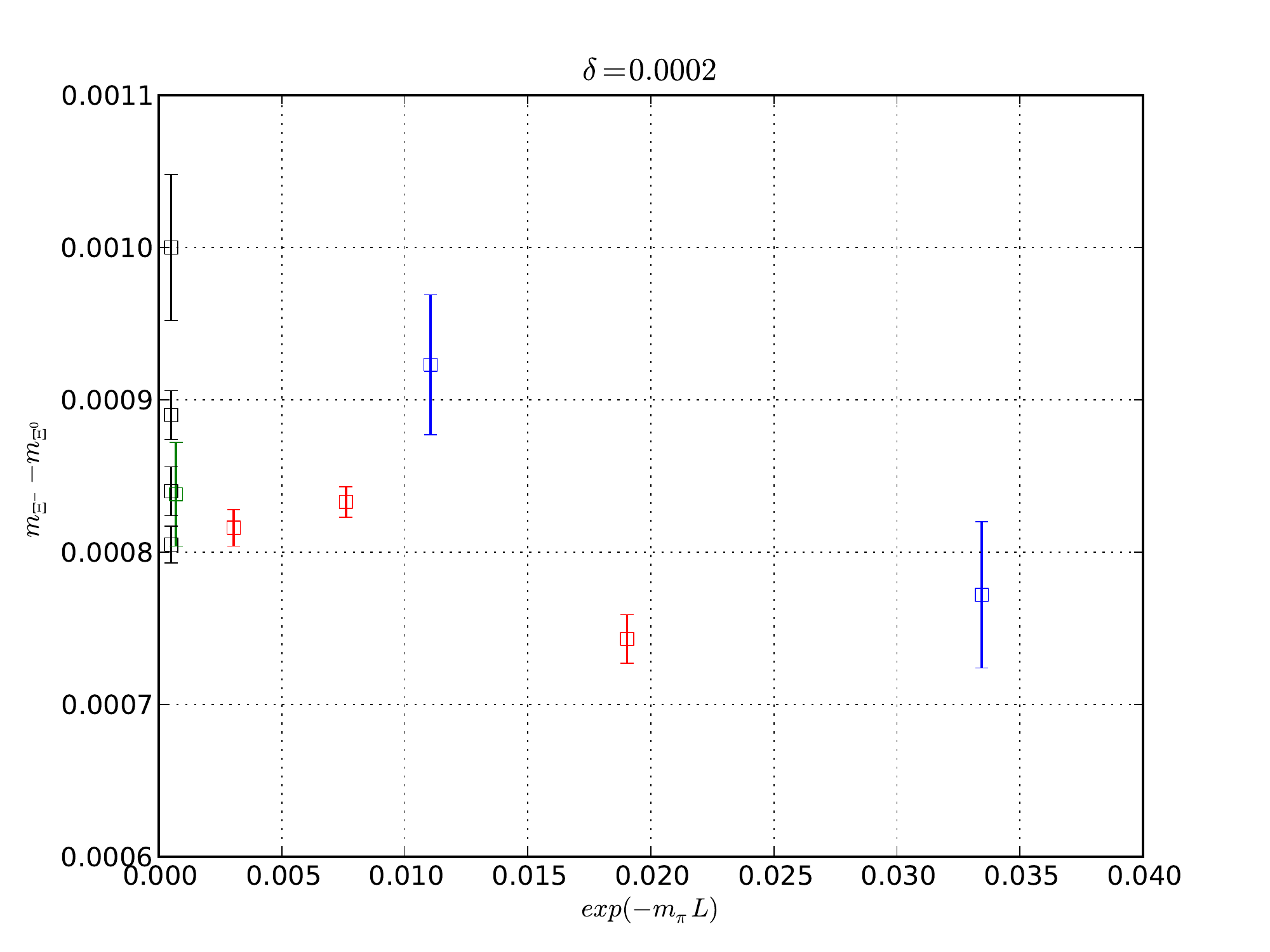}
&
\includegraphics[width=0.5\textwidth]{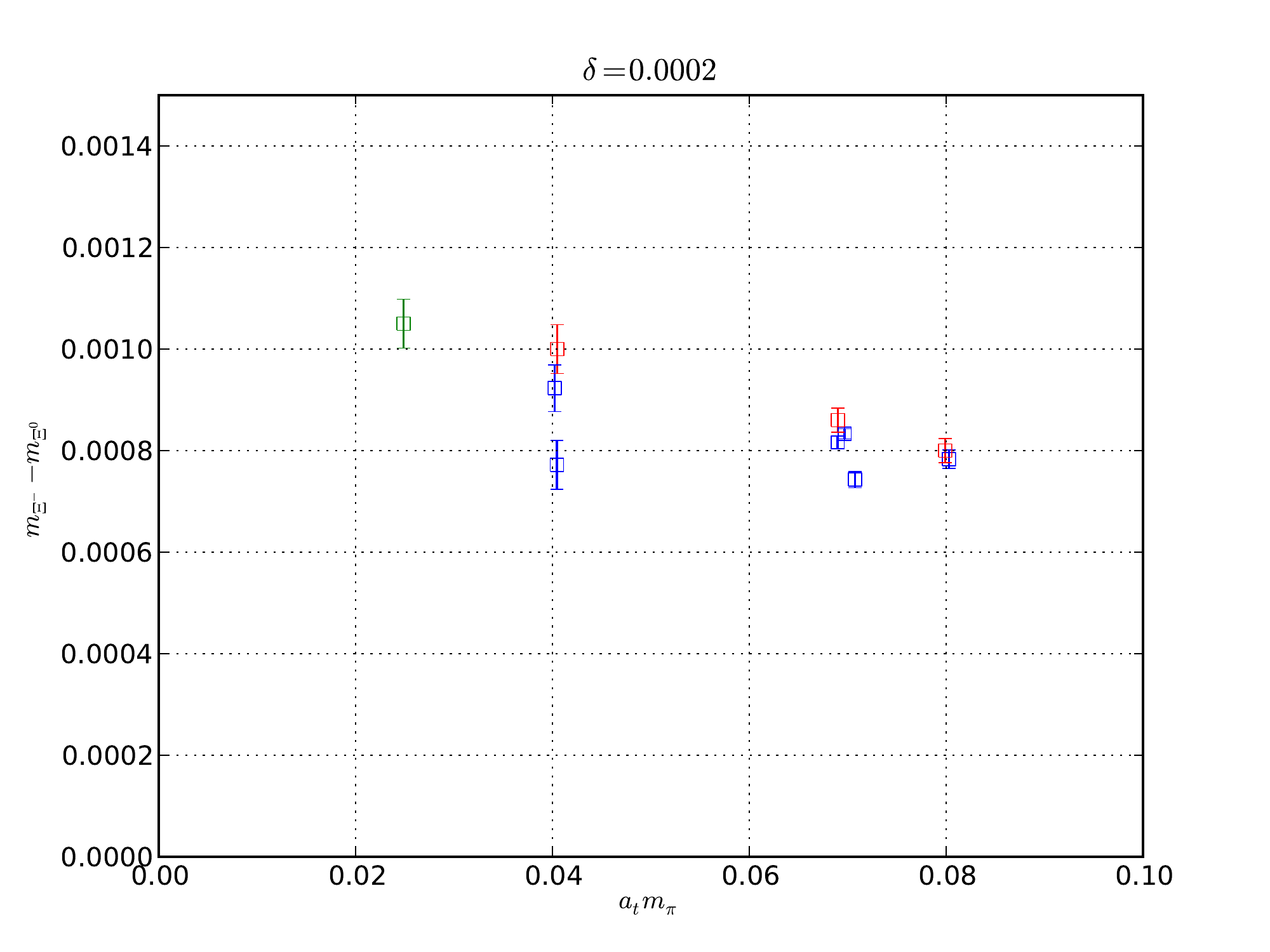}
\\
(a) & (b)
\end{tabular}
\caption{(a) plots the volume dependence for the $\Xi$ splitting at $a_t \d = 0.0002$, as well as a simple volume extrapolation.  (b) plots the pion mass dependence of the splittings.  The blue points are the data while the red points are volume extrapolated.  The green point is extrapolated in $m_\pi$ and volume.\label{fig:VmpiDependence}}
\end{figure}

In Fig.~\ref{fig:xiRatio}, we display effective mass plots of the ratio correlators $\Xi^-(t) / \Xi^0(t)$ from one of our ensembles.  In the left plot, we display the effective masses, demonstrating the precision of our calculation.  In the right plot, we have scaled these effective masses by $0.001 / \d$, demonstrating that for the lightest three values of $\d$, there is very little indication of non-linear in $\d$ behavior of the mass splittings.
In Table~\ref{tab:XiMeV}, we list our determination of the $\Xi$ mass splittings on the various ensembles, using our scale setting to convert to \texttt{MeV}.
\begin{table}[bt]
\begin{tabular}{cccc|cc|ccccc}
\hline\hline
\multicolumn{4}{c|}{ensemble: $a_t m_s = -0.0743$}& $m_\pi$& $m_K$& $\d = $& 0.0002& 0.0004& 0.0010& 0.0020 \\
\hline
$L$& $T$& $a_t m_l$& $a_t m_l^{val}$& $[\texttt{MeV}]$& $[\texttt{MeV}]$& \multicolumn{5}{c}{$m_{\Xi^-} - m_{\Xi^0}$~\texttt{[MeV]} } \\
\hline
16& 128& -0.0830& -0.0830& 490& 630&& 4.8(1)&--&23.2(3)& -- \\
16& 128& -0.0840& -0.0840& 420& 592&& 4.5(1)& 9.0(2)& 22.7(5)& 46.8(1.7) \\
20& 128& -0.0840& -0.0840& 420& 592&& 5.1(1)& --& --& --\\
24& 128& -0.0840& -0.0840& 420& 592&& 4.9(1)& 10.1(2)& 25.0(3)& --\\
24& 128& -0.0860& -0.0860& 244& 506&& 4.7(3)& --& --& --\\
32& 256& -0.0860& -0.0860& 244& 506&& 5.6(3)& --& --& --\\
\hline\hline
\end{tabular}
\caption{\label{tab:XiMeV} Resulting $\Xi$ splittings in our calculation.}
\end{table}
We unexpectedly found a significant volume dependence in our results.  In Fig.~\ref{fig:VmpiDependence}(a), we plot results of the $\Xi$-splitting for $\a_t \d = 0.0002$ for two masses on different volumes \textit{vs} $\textrm{exp}(-m_\pi L)$.  In Fig.~\ref{fig:VmpiDependence}(b), we plot these same results \textit{vs} $m_\pi$.  One sees the volume dependence is at least as significant as the pion mass dependence.
To perform the extrapolation, we use the NNLO formula, Eq.~\eqref{eq:MsplitNNLO}
including finite volume corrections, giving us the determination
\begin{equation}
Z_m^{latt} \d_{phys} = 0.000216(15)(\textrm{sys.})(13)\texttt{[ l.u.]} \quad \textrm{(preliminary)}\, .
\end{equation}
The first uncertainty is statistical, the second unquoted uncertainty is a fitting systematic and the third and dominant uncertainty is from the quoted EM self-energy corrections of Gasser and Leutwyler.
Using our lattice results for the nucleon masses, this allows us to predict the strong contribution to the splitting
\begin{equation}
m_n - m_p = 3.40 \pm 0.20 \pm \textrm{ sys.} \pm 0.74 \texttt{ [MeV]}\, ,
\end{equation}
which is comparable to the values of Refs.~\cite{Beane:2006fk,Blum:2010ym}.
These results make use of the EM self-energy corrections of Ref.~\cite{Gasser:1982ap} which did not account for the subtraction term known to be needed~\cite{Harari:1966mu}.

%
\section{Conclusions\label{sec:conclusions}}
In addition to having different lattice computations of given quantities to test for universality, it is also important to have different approaches to check for consistency.  
We have presented an independent means of determining the light quark mass isospin breaking parameter, utilizing the baryon spectrum.  
We have used the prudent choice of splitting the valence quark masses symmetrically about the degenerate sea quark mass, a choice which mitigates the partial quenching artifacts~\cite{WalkerLoud:2009nf}.
In order for this method to be successful, the electromagnetic self energy of the nucleon must be addressed, an old but unresolved issue.  If a means of determining the required subtraction constant can not be found, then a second consistency check can still be performed; the isospin breaking parameter can be determined from the meson spectrum, and then used with a lattice calculation of the strong contribution of the neutron-proton mass splitting to indirectly compute this subtraction constant.  These results can then be used to post-dict the mass splitting $m_{\Xi^-} - m_{\Xi^0}$.  In either case, the phenomenologically interesting dependence of the nucleon mass splitting on $\d$ can be determined.

\acknowledgments
This project utilized the Chroma software suite~\cite{Edwards:2004sx}.  The computations were performed on the \textit{Sporades} cluster at the College of William and Mary as well as the Thomas Jefferson National Accelerator Facility with time awarded by the USQCD Collaboration.  We thank the Hadron Spectrum Collaboration for use of their gauge configurations.  The work of AWL was supported in part by the Jeffress Memorial Trust, grant J-813 and the DOE OJI grant DE-FG02-07ER41527.

\end{document}